# Correlating Photovoltaic Soiling Losses to Waveband and Single-Value Transmittance Measurements

Leonardo Micheli,[1,2,*] Jose A. Caballero,[2] Eduardo F. Fernandez,[2,**]
Greg P. Smestad,[3] Gustavo Nofuentes,[2] Tapas Mallick,[4] and Florencia Almonacid[2]

[1] *National Renewable Energy Laboratory, Golden (CO), USA*
[2] *University of Jaén, Jaén, Spain*
[3] *Sol Ideas Technology Development, San José (CA), USA*
[4] *University of Exeter, Penryn, UK*

*Leonardo Micheli. Tel: +34 953211744, E-mail: lmicheli@ujaen.es
**Eduardo F. Fernández. Tel: +34 953213520, E-mail: eduardo.fernandez@ujaen.es

**Abstract**[*]
This paper presents the results of an investigation on the spectral losses of photovoltaic (PV) soiling. The transmittance of a glass coupon exposed to natural soiling outdoors in Jaén, southern Spain, has been measured weekly and used to estimate the soiling losses that various types of photovoltaic materials would experience if installed in the same location. The results suggest that measuring the hemispherical transmittance of the soiling accumulated on a PV glass coupon can give enough information to quantify the impact of soiling on energy production. Each PV technology is found to have a preferred spectral region, or a specific single wavelength, for which the transmittance through a PV glass coupon could be used for the best estimation of soiling losses. Overall, considering the average spectral transmittance between the extreme wavelengths of the material-specific absorption band, or the transmittance of soiling at a single wavelength between 500 and 600 nm yields the best estimations for different PV technologies. The results of this work can lead to innovative approaches to detect soiling in the field and to estimate the impact of spectral changes induced by soiling on PV energy production.

---

[*] **Abbreviations**
$A_{PV}$: active area of the photovoltaic device
a-Si: amorphous silicon
AST: average soiling transmittance
CdTe: cadmium telluride
CIGS: copper indium gallium diselenide
EG: spectral distribution of irradiance on the photovoltaic surface
Isc: short-circuit current
MAPE: mean absolute percentage error
MPE: mean percentage error
m-Si: monocrystalline silicon
NIR: near infrared
p-Si: polycrystalline silicon
$R^2$: coefficient of determination
$r_s$: soiling ratio
SR: spectral response of the unsoiled PV module
UV: ultra-violet
VIS: visible
τ: hemispherical spectral transmittance

**Graphical Abstract**

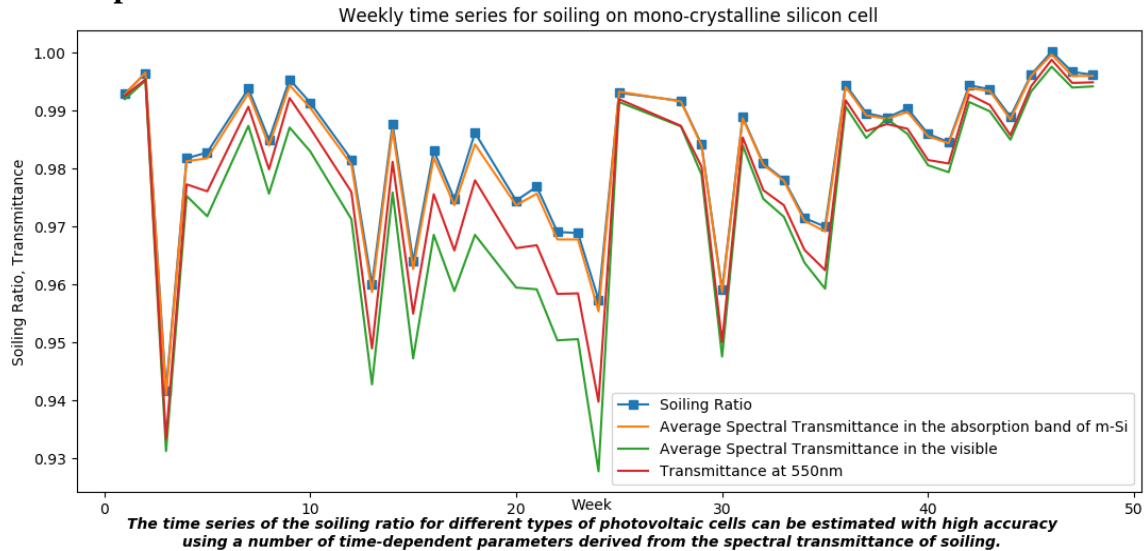

*The time series of the soiling ratio for different types of photovoltaic cells can be estimated with high accuracy using a number of time-dependent parameters derived from the spectral transmittance of soiling.*

**Highlights**

- The transmittance of an outdoor-mounted PV glass coupon was monitored for a year
- Broadband or single-wavelength transmittance can predict the PV panel soiling ratio
- The best single-wavelength estimations are found between 500 and 600 nm
- The average transmittance over the PV material-specific band has a high correlation
- The average transmittance in the visible correlates better than the UV or NIR bands

**1. Introduction**

The accumulation of dust, particles, and dirt on the surface of photovoltaic (PV) modules reduces the intensity of the light transmitted through the cover glass—and, therefore, the amount of energy generated by the solar cells. This issue, known as *soiling*, affects PV systems worldwide, causing power losses as high as 70% in the worst scenarios [1]. These losses are due to the drop in optical transmittance, because soiling absorbs part of the incoming sunlight and increases the fraction of reflected light. Soiling can be mitigated by cleaning the PV modules, when the energy losses are higher than the cleaning costs. This means that an accurate soiling monitoring system is required to properly address this issue, thereby increasing the energy yield while minimizing the operating and maintenance cost.

It is important to highlight that, along with the broadband reduction in irradiance, soiling also changes the spectrum of the transmitted light, causing larger transmittance drops in the blue region, meaning that PV technologies with different bandgaps can be differently affected by soiling [2–5]. Hence, measuring and analyzing soiling should not disregard the spectral effects that soiling can have on various PV technologies; but determining the spectral component of soiling on fielded PV modules is still challenging.

The current soiling monitoring technologies calculate losses by comparing the performance of a soiled device with the performance of a device that has been kept clean throughout the time period. This means that they can calculate losses occurring for only a single PV technology, unless multiple soiled devices are employed. Different and more affordable solutions are required to estimate the impact of soiling on the various PV technologies. Therefore, it is essential to investigate the correlations connected to the spectral nature of natural soiling, the irradiance spectra, and the soiling losses through extended experimental campaigns.

In light of the above, the scope of this work is to analyze the spectral components of the soiling losses, and to investigate the impact of soiling on the various wavelengths of the irradiance spectrum and on the various PV technologies. This study shows how transmittance measurements can be used to estimate the soiling losses of PV modules and also how a single wavelength measurement can be used to estimate the soiling losses of various PV technologies. The results of this work are expected to contribute identifying innovative solutions and metrics to measure the soiling ratio of any PV technology, lowering the cost of soiling monitoring and limiting its impact on the PV cost competiveness.

## 2. Background

Soiling is the result of the interaction of a number of environmental factors [6–12]. Typically, the soiling of a PV system is monitored by using soiling stations, where the electrical output of a soiled PV device (*soiled device*) is compared with the output of the PV device under clean conditions (*control device*). Generally, two PV cells or modules are employed: one of them is regularly cleaned (*control device*), whereas the second is left to soil naturally (*soiled device*) [13]. Despite the simple approach, these stations require regular cleanings, which might be expensive to perform, in order to limit the uncertainty of the measurement, as the error associated with stations that are not well maintained can be as high as the soiling loss [14].

In order to eliminate the requirement for washing a reference device and to lower the cost of soiling monitoring, new soiling detection technologies have been recently developed. A first product, called DustIQ and developed by Kipp&Zonen, uses a photodiode to measure the backward reflection of a soiled PV glass, mounted next to or within a PV array [15]. A second product, named MARS and developed by Atonometrics, uses a camera to analyze the impact of soiling on a soiled PV glass, by measuring the brightness of the pixels in the camera's field of view [16]. These two products do not require any cleaning, and measure the soiling accumulated on PV glass to estimate the soiling losses occurring on similarly mounted PV modules located nearby. In other geometries, in contrast, only one PV cell is used as both the soiled and the control device [17]. In this case, a cover glass is placed on top of the cell to collect soiling and is removed only to take the control device measurement.

Along with their dependence on the broadband irradiance, the performance of PV systems have also a dependence on the spectral content of the sunlight, as well as on the specific spectral absorption band of the technology [18]. Soiling has also an effect on the spectrum of the irradiance hitting the semi-conductive material and therefore its impact varies depending on the irradiance spectrum and on the spectral response of the PV technology. Despite that, none of the current soiling monitoring technologies is able to estimate the spectral component of the soiling losses and, therefore, to correct the soiling measurement for different PV technologies.

So far, most of the research addressing the spectral impact of soiling has been conducted by analyzing artificial soiling [2,19,20]. These studies have been crucial to understanding the effects of soiling on the performance of PV systems. But even if artificial soiling makes it possible to conduct the investigation in a controlled environment, it limits the understanding of the spectral nature of soiling in actual outdoor conditions, where types of soiling and deposition rates are different and can vary with time. The first results based on an outdoor experimental campaign on spectral soiling losses, where natural soiling was collected on glass coupons exposed for two-months in eight locations worldwide, were recently presented [3]. The results of that work confirmed the higher attenuation of soiling at lower wavelengths, as previously indicated by other studies [2,4]. Despite all these important efforts, the relation between the spectral nature of soiling and its impact on each PV technology needs to be further investigated because their different absorption bands can lead to dissimilar losses among the various PV materials even in the same conditions of soiling. In this light, the current work focuses on understanding the spectral profile of natural soiling and its impact on various PV technologies, and on identifying the correlations between the transmittance of soiling and the losses in PV. These correlations can make it possible to estimate the soiling losses of various PV technologies by measuring the transmittance of soiling within a limited waveband or even at a single wavelength, opening the possibility to innovative soiling detection mechanisms.

In order to measure the spectral transmittance of soiling and to calculate its impact on the soiling losses of PV modules, a 4 cm × 4 cm PV glass coupon has been outdoor mounted and monitored in this study. The use of glass coupons to analyze and monitor the soiling of PV modules is a standard method for the soiling community, already employed for both research and commercial purposes. Outdoor mounted PV coupons were used by Burton et al. [21] to study the composition and the size distribution of soiling deposited on PV modules. The same setup, replicated also in a second location, was used by Boyle et al. [22] to investigate the deposition of particle matter on PV systems. Similarly, the dust accumulated on PV glass coupons was used by Conceição et al. [23] to estimate the soiling loss occurring on two mono-Si PV modules nearby. A team at NREL exposed PV glass coupons for one year in five different locations to understand the effectiveness of anti-soiling coatings and to quantify the abrasion due to external agents and various cleaning methods [24,25]. Nayshevsky et al. [26] used glass coupons to investigate the use of hybrid hydrophobic-hydrophilic coatings to improve the collection of dew and so to decrease the soiling rates on PV modules. Also, as mentioned, novel maintenance-free soiling detectors quantify the soiling accumulated on a PV glass to estimate the soiling losses of PV modules [15,16]. Gostein et al. [27] demonstrated an high correlation between the soiling losses estimated by MARS and the losses measured on a PV cell, both covered by PV glass coupons artificially deposited with three different types of dusts were placed on top of them. Korevaar et al. [28] compared the measurements of DustIQ with the soiling ratio measured by a soiling station in Morocco, showing a good correlation between the two profiles over a short period.

Despite the wide usage of coupons for PV soiling studies, further studies are already being conducted to understand the different soiling mechanisms occurring on stand-alone glass vs. PV modules [29]. The present paper uses the transmittance measured on a PV glass to estimate the soiling losses that a module exposed to the same soiling conditions would experience. Therefore, this study does not directly compare two different measurements as the transmittance of a PV glass and the power loss of a PV module. For this reason, the discussion on the different soiling deposition mechanisms occurring on glass and modules is considered out of the scope of this work.

### 3. Materials and Methods

#### *3.1. Soiling spectral indices*

The most commonly used metric to quantify the impact of soiling is the Soiling Ratio ($r_s$), which expresses the ratio between the output of a soiled PV device and the output of the same device under clean conditions [30]. Similar to the transmittance, the higher the soiling ratio, the less the soiling deposited on the modules. The soiling ratio assumes a value of 1 in clean conditions, with no soiling, and decreases while soiling deposits. The soiling losses, expressed as the fractional loss in power due to soiling, can be estimated as: 1- $r_s$. The short-circuit current can be used as the electrical output for the calculation of the soiling ratio if soiling is uniform, whereas the maximum power point is required for a better estimate of nonuniform soiling [9,13]. In this work, the nonuniform effects of soiling are not considered: therefore, the instantaneous soiling ratio at any time *t* is calculated as follows:

$$r_s(t) = \frac{Isc_{soil}(t)}{Isc_{ref}(t)} \quad (1)$$

where $Isc_{soil}$ and $Isc_{ref}$ are the short-circuit currents of a soiled PV device and of the control device, respectively. Considering this, the soiling ratio for a specific period of time T can be obtained as the average of the measured instantaneous soiling ratios by the following expression:

$$\overline{r_s}(T) = \frac{1}{n}\sum_{i=1}^{n} r_s(t) \quad (2)$$

where n is the number of measurements over the period of time T.

The time-dependent short-circuit currents of Equation (1), $Isc_{soil}$ and $Isc_{ref}$, could be obtained either experimentally from two monitored PV devices, or, as in the present work, calculated by solving the following expressions [31]:

$$Isc_{ref}(t) = A_{PV} \int_{\lambda_1}^{\lambda_2} E_G(\lambda, t) SR(\lambda) d\lambda \quad (3)$$

$$Isc_{soil}(t) = A_{PV} \int_{\lambda_1}^{\lambda_2} E_G(\lambda, t) \tau_{soiling}(\lambda, t) SR(\lambda) d\lambda \quad (4)$$

where $\lambda_1$ and $\lambda_2$ are the lower and upper limits of the absorption band of each PV device's absorber material (i.e., m-Si, CdTe, CIGS), $A_{PV}$ is its active area, $SR(\lambda)$ is the spectral response, $\tau_{soil}(\lambda,t)$ is the hemispherical spectral transmittance of soiling accumulated on the surface of the soiled device, and $E_G(\lambda,t)$ is the actual spectral distribution of the solar irradiance on the plane of the PV panels.

At the same time, the average spectral transmittance (AST) of soiling across a specific spectral waveband can be calculated with the following relationship:

$$AST_i(t) = \frac{1}{\lambda_{2_i} - \lambda_{1_i}} \int_{\lambda_{1_i}}^{\lambda_{2_i}} \tau_{soiling}(\lambda, t) d\lambda \quad (5)$$

where $\lambda_{1i}$ and $\lambda_{2i}$, respectively, are the shortest and longest wavelengths in the selected waveband *i*. The various wavebands considered in this work describe either a specific spectral region or an individual PV material's absorption band and are listed in Table 1. Note that the lower and upper limits of the spectral region bands are defined by considering the absorption bands of the PV materials; the shortest and longest wavelengths selected are, respectively, 300 nm for the ultraviolet and 1,240 nm for the near-infrared regions.

Table 1. Wavebands considered in the present study.

| Waveband | | $\lambda_1$ [nm] | $\lambda_2$ [nm] |
|---|---|---|---|
| Spectral regions | Ultraviolet (UV) | 300 | 400 |
| | Visible (VIS) | 400 | 700 |
| | Near-infrared (NIR) | 700 | 1,240 |
| PV material absorption bands | Monocrystalline silicon (m-Si) | 340 | 1,190 |
| | Polycrystalline silicon (p-Si) | 310 | 1,180 |
| | Amorphous silicon (a-Si) | 300 | 790 |
| | Cadmium telluride (CdTe) | 310 | 880 |
| | Copper indium gallium diselenide (CIGS) | 370 | 1,240 |
| | Perovskite | 300 | 820 |

*3.2. Experimental campaign*

### 3.2.1 Location

A 48-week experiment—from January 2017 to January 2018—was conducted on the roof of the A3-building at the University of Jaén, in Jaén, Spain (latitude 37º49'N, longitude 3º48'W, elev. 457 m). Jaén is a medium-size town located in southern Spain with a high annual energy resource (more than 1,800 kWh/m$^2$), and extreme temperatures ranging from less than 5ºC in winter to more than 40ºC in summer [32]. Atmospheric conditions are described by low-medium values of precipitable water, turbidity, and airborne particulate matter, even if this can periodically reach unusually high values due to specific and stochastic events such as Saharan dust storms or the burning of branches from olive groves in the local region [33].

### 3.2.2 Transmittance

One Diamant® low-iron glass coupon 4 cm × 4 cm in size and 3 mm thick from Saint-Gobain Glass was placed horizontally outdoors to capture natural dust (*Soiled Coupon*). The same mounting configuration shown in [3] has been used. The coupon was left to soil naturally, and was manually cleaned only once after 24 weeks, at the half way point of the experimental period, to check for the absence of permanent degradation on the glass. Its hemispherical transmittance was measured weekly within a wavelength range between 300 and 1,240 nm, using a Lambda 950 spectrophotometer with a 60-mm-diameter integrating sphere at the Center of Scientific-Technical Instrumentation (CICT) of the University of Jaén. The general arrangement is outlined in Figure 1. Another sample (*Control Coupon*) was stored in a dust-free box to prevent its optical transmittance characteristics from being adversely affected from accidental soiling, and it was used as the baseline for each measurement. Its transmittance also allowed us to check the quality and repeatability of weekly measurements. The soiling transmittance is obtained from the measurement as follows:

$$\tau_{soiling}(\lambda) = \frac{\tau_{soil}(\lambda)}{\tau_{ref}(\lambda)} \quad (6)$$

where $\tau_{soil}(\lambda)$ and $\tau_{ref}(\lambda)$ are the spectral transmittance of the Soiled Coupon and Control Coupon, respectively, for one of the wavelength ranges described in Table 1.

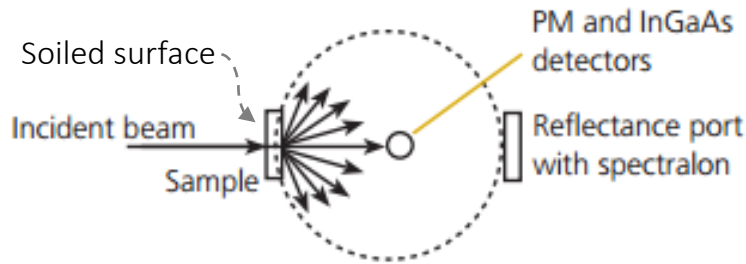

Figure 1. Schematic of the hemispherical transmittance measurement using the integrating sphere. The detectors are mounted orthogonally from the beam direction on the exterior of the sphere.

### 3.2.3 Irradiance

The global spectral irradiance between the 350-nm and 1,050-nm wavelength band was measured at 5-minute intervals using a weatherproof spectroradiometer (EKO® MS700) oriented south and tilted 30º above the horizontal. This angle has been selected to maximize the collected irradiation over the year [34]. This instrument presents a spectral resolution of 10 nm and a temperature dependency within ±1% for temperatures ranging from -20ºC to +50ºC. The expanded uncertainties of the instrument are ±10.90%, ±4.20%, and ±4.10%, respectively, for the 350–450-nm, 450–900-nm, and 900–1,050-nm wavebands, according to the certificate of calibration provided by the manufacturer. The absorption bands of some PV devices (see Table 1 and Figure 2) go beyond the measurement range of the spectroradiometer. This limitation has been overcome by using the methodology proposed by Martín and Ruiz [35]. The missing wavebands have been estimated by scaling the AM1.5G reference spectrum according to the ratio between the integrated actual and referenced spectral irradiance in the range of 700-1,050 nm. This methodology was used and validated by Nofuentes et al. [36] to elucidate the impact of the average photon energy (APE) on the spectral mismatch factor (MM). In addition, all measurements recorded at irradiance levels below 300 W/m$^2$ have not been taken into account so as to avoid the non-linear performance of PV cells at such low irradiance values [36–38]. Nevertheless, these low irradiance levels do not play a significant role in the annual electrical output of PV systems at locations with a high-energy solar resource, such as the location considered in this study [39–41]. In addition, all the measurements with an incident angle equal to or greater than 60$^0$ have been removed to reduce the impact of the increased glass Fresnel reflection [42]. This approach also automatically excludes conditions in which the impact of soiling has been found to be strongly related to the angle of incidence [43].

### *3.3. Methodology*

PV materials have different spectral absorption bands and different spectral responses, as shown in Figure 2 for the six PV technologies investigated in this study. In addition, the transmittance of soiling has a nonuniform spectral distribution, with higher losses at shorter wavelengths, as shown in the bottom chart of Figure 3. This means that soiling can have different impact on the various PV materials, which is a result of the product of the spectral response with the time-dependent irradiance and soiling transmittance spectra. Note that discussing the effects of soiling on the different PV technologies is outside the scope of this paper, which only focuses on the relationship between soiling losses and soiling transmittance. With this in mind, we have established the following procedure to conduct the analysis presented in this paper:
1. Measuring the spectral transmittance of soiling ($\tau_{\text{soiling}}(\lambda)$) collected on the Soiled Coupon once per week by using Equation (6).

2. Calculating the soiling ratios by using Equations (1), (2), (3), and (4). The SR of each PV device, the $\tau_{soiling}(\lambda)$ obtained in step 1, and the irradiance spectra recorded during the same day are used as inputs.
3. Estimating the average transmittance of soiling (AST($\lambda$)) for the regions of the spectrum and the PV devices listed in Table 1 using Equation (5) and $\tau_{soiling}(\lambda)$.
4. Comparing the soiling ratios obtained in step 2 for the different wavebands investigated in step 3 by using several standard statistical metrics.

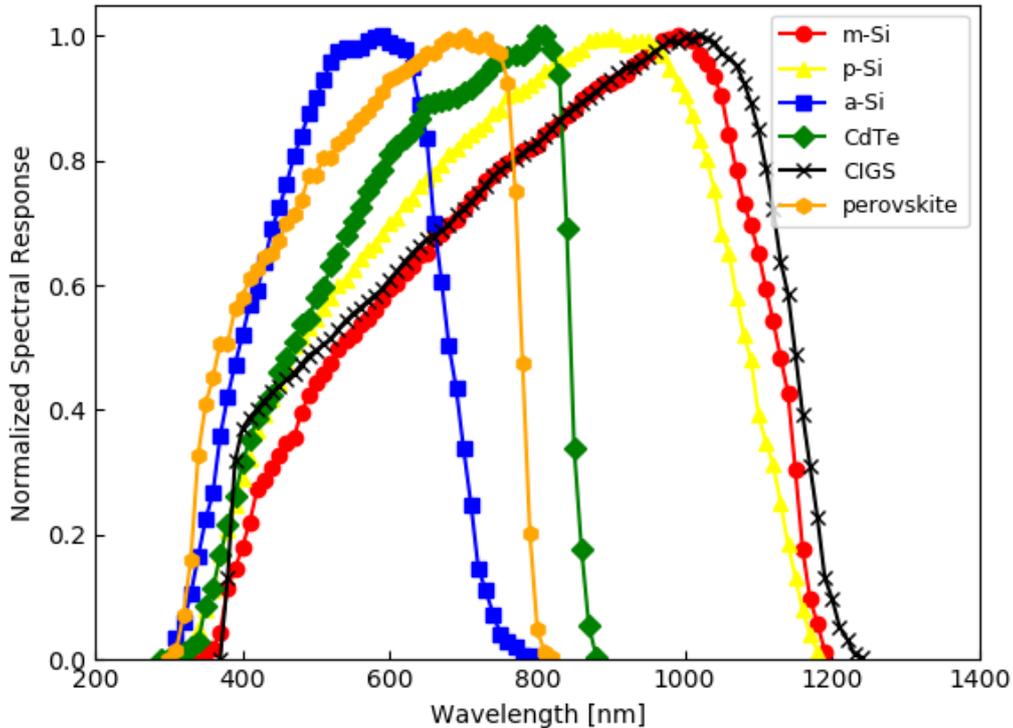

Figure 2. Normalized spectral response of the six PV materials considered in this study.

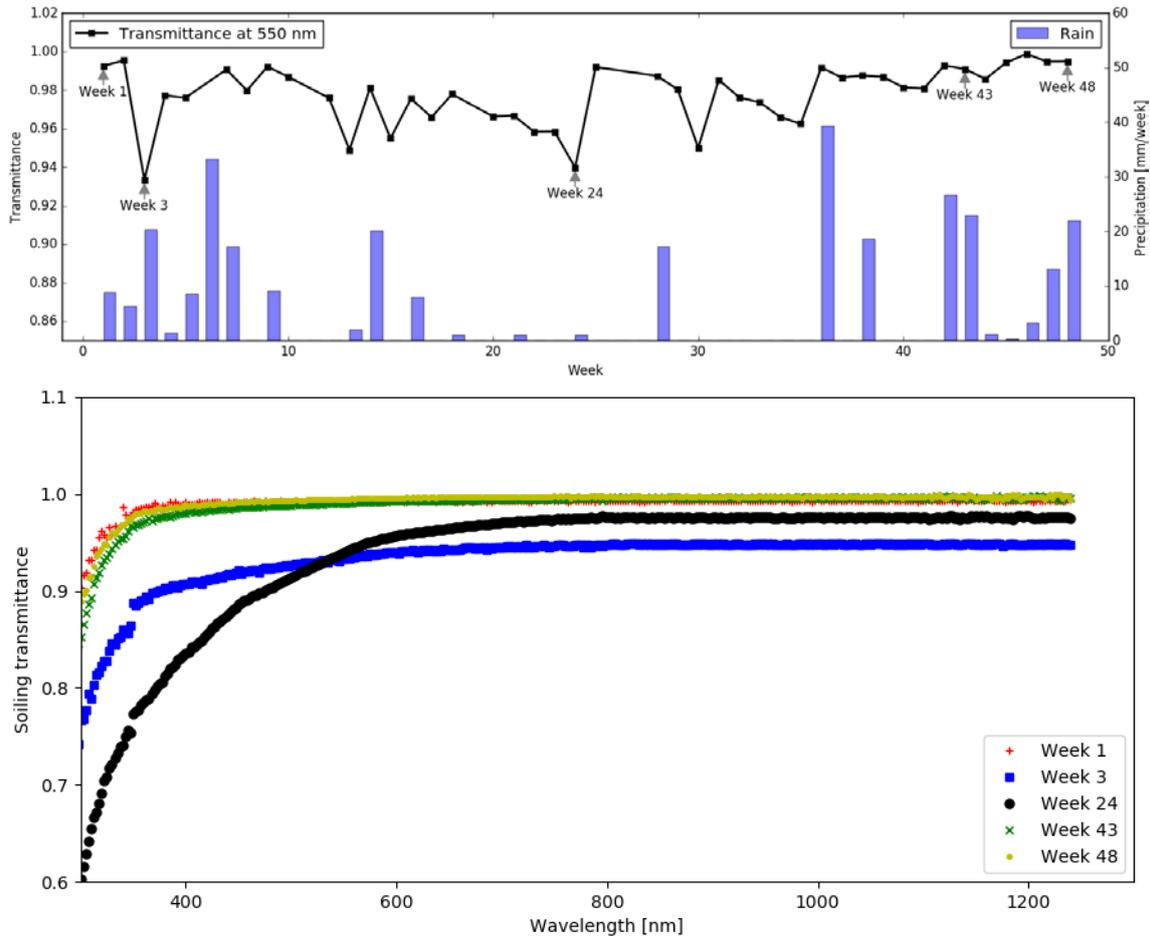

Figure 3. Top: weekly transmittance of the coupon at 550 nm (left y-axis) and accumulated precipitation (right y-axis) during the investigation period. Bottom: soiling transmittance for five representative weeks of the data collection period.

*3.4. Data Collection*

Soiling tends to accumulate during dry periods. However, it can be washed away by rainfall [8]. The time series of the transmittance at 550 nm of the coupon exposed for one year in Jaén is shown in Figure 3 (top chart). Five representative weeks are labeled, and the spectral transmittance during those are shown in the lower plot. In order to prove the relations between rainfall and soiling, the weekly accumulated precipitation, recorded by an atmospheric station MTD 3000 from Geonica S.A. located on the rooftop of the one of the buildings of the University of Jaén, is also reported on the right y-axis of the top chart.

The transmittance decreases between week 1 and week 3, due to an extreme soiling event, recorded also by the local particulate matter station of the Andalucía air quality monitoring and control network located around 1 km from the experimental set-up, which recorded a $PM_{10}$ concentration of 87.5 μg/m$^3$, out of an annual average of 25.8 μg/m$^3$. The transmittance then raises in week 4 thanks to the large amount of precipitation recorded in that week that has a cleaning effect on the coupon. After few soiling and cleaning events between weeks 10 and weeks 20 that keep limited soiling on the coupon, the transmittance reaches a minimum value in week 24. After week 24, the soiled glass is manually cleaned to check for signs of degradation; as shown at the top of Figure 3, its transmittance is restored to 1, proving no degradation due to external exposure compared to Coupon 0. A clear soiling trend can also be seen between weeks 31 and 35, during which the transmittance decreases at a rate of -0.0056 per week ($R^2 = 0.97$), down to a minimum value of 0.962. During the last weeks of the data collection, the transmittance is quite high and consistent because of the frequent rainfall events that keep the coupon clean.

This paper considers only the hemispherical transmittance of soiling. Even if less affected by soiling than the direct component [3], the hemispherical transmittance has been preferred because it is more representative of the actual effect of soiling on power conversion. Photovoltaic modules, indeed, can convert both the direct and diffuse components of the light, as well as the part that is scattered by soiling. The hemispherical transmittance measurement can capture all these components. Therefore, any transmittance mentioned in the document must be considered as hemispherical.

## 4. Results and Discussion

In this section, two different analyses are carried out. First, the soiling ratio is estimated for three different spectral regions and for a region specific to the spectral response band of each PV material. Second, an estimation is made by using a single wavelength to facilitate the quantification of the spectral impact of soiling as accurately as possible with a simple single measurement. Each analysis is conducted by using different statistical indexes: the determination coefficient ($R^2$), the mean absolute percentage error (MAPE), and the mean percentage error (MPE). These magnitudes have been calculated by means of the following expressions [44]:

$$R^2 = \left( \frac{\sum_{i=1}^{n}(r_s - \overline{r_s})(Z - \bar{Z})}{\sqrt{\sum_{i=1}^{n}(r_s - \overline{r_s})^2 \sum_{i=1}^{n}(Z - \bar{Z})^2}} \right)^2 \quad (7)$$

$$MAPE\ (\%) = \frac{100}{n} \sum_{i=1}^{n} \left| \frac{Z - r_s}{r_s} \right| \quad (8)$$

$$MPE\ (\%) = \frac{100}{n} \sum_{i=1}^{n} \frac{Z - r_s}{r_s} \quad (9)$$

where n is the number of soiling ratio data points and Z represents the soiling ratio predicted through the average spectral transmittance or a single wavelength transmittance data points used to estimate the soiling ratio. The coefficient of determination measures the quality of the fit between the soiling ratios and the Z values. It has a value of 1 if the Z points predict the soiling ratios with a linear equation with no error, and it has a value of 0 if no linear correlation exists between the soiling ratios and the Z points. The MAPE measures the average value of the absolute errors between the soiling ratios and their calculated values (Z points). It has a value of 0 if the soiling ratios and the Z value are the same, and it increases depending on the number and the magnitude of the errors in the prediction. The MPE is a metric calculated similarly to MAPE, but it takes into account the actual values of the errors instead of their absolute values, and it gives information on any systematic bias in the prediction: it is positive if the predicted values tend to overestimate soiling; otherwise, it is negative.

### *4.1. Analysis of spectral waveband*

In this subsection, we investigate the correlations between the average spectral transmittance across different spectral bands and the soiling ratio. In Figure 4, the soiling ratio, calculated weekly using Eq. (2), is plotted against the average spectral transmittance of the ultraviolet, visible, and near-infrared regions, as given by Eq. (5). The best linear fits and the coefficients of determination ($R^2$) obtained for each PV technology in each region are also reported in the charts. For better readability, only three PV materials with high (a-Si), intermediate (CdTe) and low (m-Si) energy gaps are represented. Figure 5 shows the current density of the three materials exposed to the reference AM1.5 global irradiance. The data for all technologies, inclusive of MAPE and MPE values, are reported in Table 2.

As shown, the quality of the best fit varies with both the spectral region and the PV technology. Using the UV portion of the light lowers the correlation for all the technologies, with MAPE percentages of 7% or higher. In particular, the MPE values are all found to be negative, meaning that AST returns lower values than the actual soiling ratios. This is not surprising because the UV region contributes little to the current generation in PV modules because it represents only a limited portion of the solar irradiance spectrum. Moreover, all the PV technologies have low spectral response in this region (Figure 5), whereas soiling causes dramatic transmittance drops [2,3]. The best result for the UV is found for a-Si because this is the technology with the highest absorption at the lowest wavelengths (Figure 5).

Overall, the best results are obtained if the visible AST is considered, with the maximum $R^2$ achieved by a-Si, perovskite, and CdTe technologies; most of their absorption occurs in this region (Figure 5). All the technologies have low MAPE (<1%), with a-Si reaching values lower than 0.1%. Soiling losses of a-Si technologies can be predicted with high accuracy by measuring the visible AST only.

The low-energy bandgap materials (m-Si, p-Si, CIGS) are the only technologies to have $R^2$ above 90% in both the visible and near-infrared regions. This is because the solar irradiance is high, and their spectral response is significant both in the visible and at higher wavelengths (Figure 5). Their MAPE is lower in the NIR than in the visible. On the other hand, $R^2$ drops and the MAPE increases for a-Si in the NIR because of the very limited spectral response in this region.

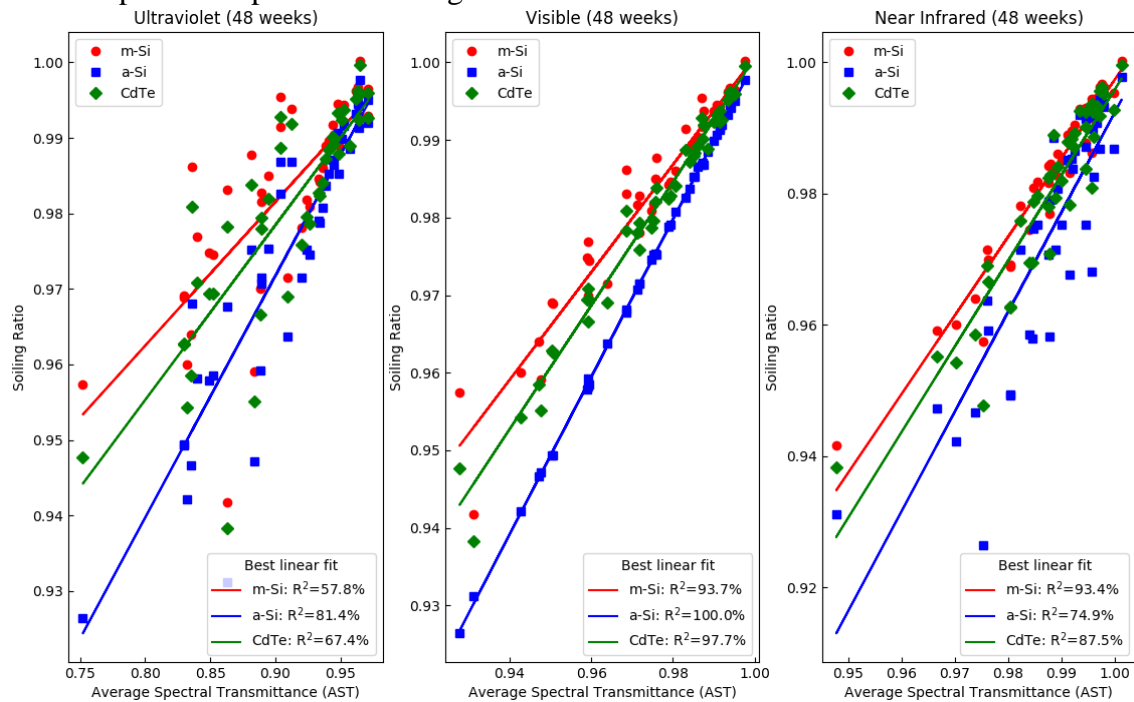

Figure 4. Soiling ratio vs AST for three representative PV materials in the three spectral regions.

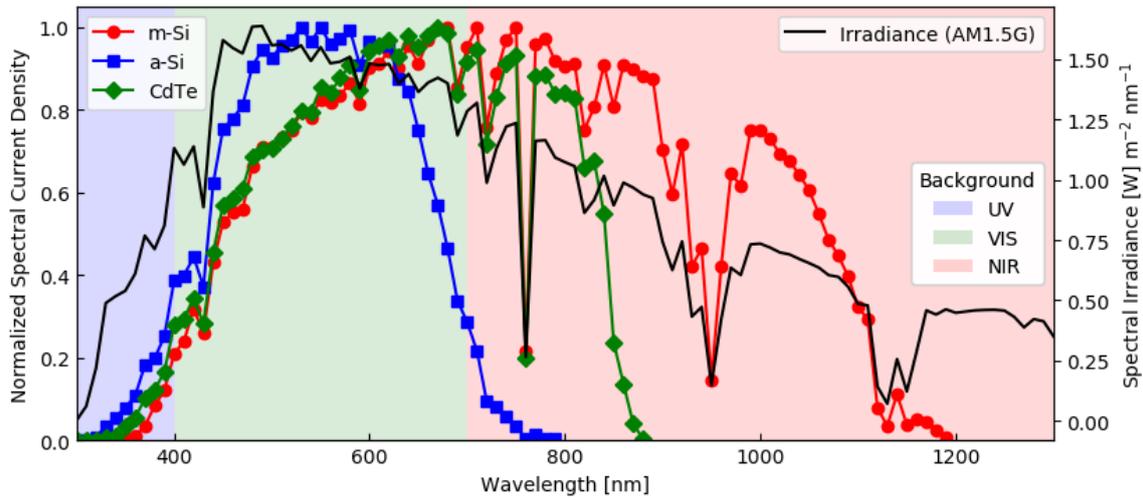

Figure 5. Current density of three PV materials under the standard AM1.5 global irradiance (ASTM G173 – 03) on the left y-axis. The irradiance spectrum (black line) is plotted on the right y-axis. The colors in the background mark the three spectral regions.

Table 2. $R^2$, MAPE, and MPE, for Jaén, Spain, between soiling ratio and transmittance in three spectral regions for the six PV materials considered.

| Material | Ultraviolet (300–400 nm) | | | Visible (400–700 nm) | | | Near-Infrared (700–1,240 nm) | | |
|---|---|---|---|---|---|---|---|---|---|
| | $R^2$ (%) | MAPE (%) | MPE (%) | $R^2$ (%) | MAPE (%) | MPE (%) | $R^2$ (%) | MAPE (%) | MPE (%) |
| m-Si | 57.8 | 7.76 | -7.76 | 93.7 | 0.88 | -0.88 | 93.4 | 0.52 | 0.52 |
| p-Si | 61.3 | 7.69 | -7.69 | 95.4 | 0.80 | -0.80 | 91.5 | 0.60 | 0.60 |
| a-Si | 81.4 | 6.94 | -6.94 | 100 | 0.04 | 0.03 | 74.9 | 1.45 | 1.45 |
| CdTe | 67.4 | 7.50 | -7.50 | 97.7 | 0.59 | -0.59 | 87.5 | 0.81 | 0.81 |
| CIGS | 60.7 | 7.71 | -7.71 | 95.1 | 0.82 | -0.82 | 91.9 | 0.58 | 0.58 |
| perovskite | 77.6 | 7.15 | -7.15 | 99.8 | 0.20 | -0.20 | 79.0 | 1.21 | 1.21 |

The visible portion of the spectrum returns the best results if materials from various energy bandgaps are investigated, even if it introduces a significant negative offset (MAPE ≥ 0.8%) for low-energy gap materials. The results for some PV technologies can be enhanced by using the specific material absorption band instead of a spectral region for calculating AST (Figure 6). Indeed, $R^2$ of at least 98%, MAPE between 0.04% and 0.65%, and negative MPE up to -0.65% for all the materials (with the worst values for a-Si and perovskite) are the result by using the PV absorption bands in calculating AST. The negative bias is because the spectral response of each material slowly grows with the wavelength from UV to visible and/or the NIR, until it peaks and dramatically drops after that (Figure 5). PV technologies have limited spectral response in that wide pre-peak region, and the irradiance region has the lowest intensity in UV. On the other hand, the AST is calculated as a simple average of the waveband transmittance (see Equation 8), giving the same weight to all the wavelengths in the spectral range, independently of the spectral response and irradiance. So, the soiling-intensive short-wavelength band has a larger impact on the AST than on the actual PV modules, leading to an overestimation of the soiling (represented by lower predicted, Z, than actual soiling ratios, $r_s$).

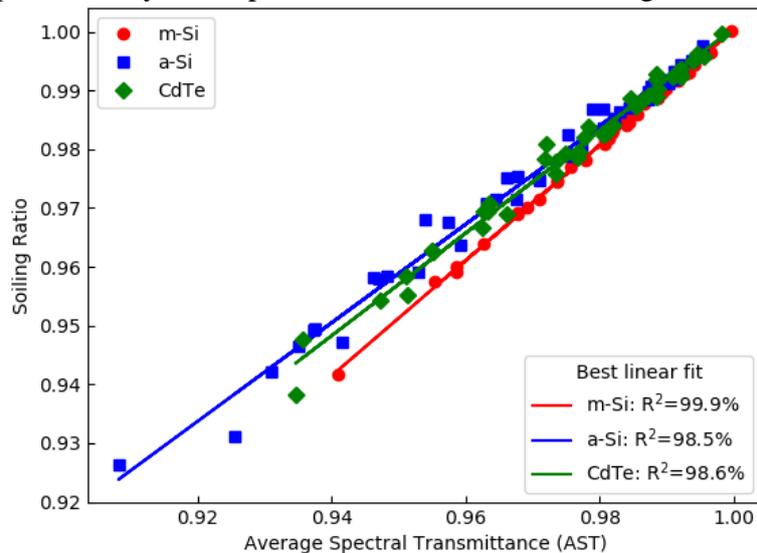

Figure 6. Soiling ratio versus AST for three representative PV material absorption bands.

*4.2. Analysis of the transmittance at single wavelength*

In the previous subsection, we showed how we use the average transmittance of a waveband to estimate the soiling losses occurring over the whole irradiance spectrum for different PV technologies. In this section, we investigate if the transmittance of a single wavelength can be used for the same purpose. For this reason, the same analysis presented earlier has been repeated using wavelengths at 50-nm steps between 300 nm and 1,000 nm. All the results are plotted in Figure 7. As can be seen, the maximum $R^2$ (≥99%) and minimum MAPE are obtained if the hemispherical transmittance at single wavelengths between 500 and 650 nm is used to estimate the soiling losses of PV materials.

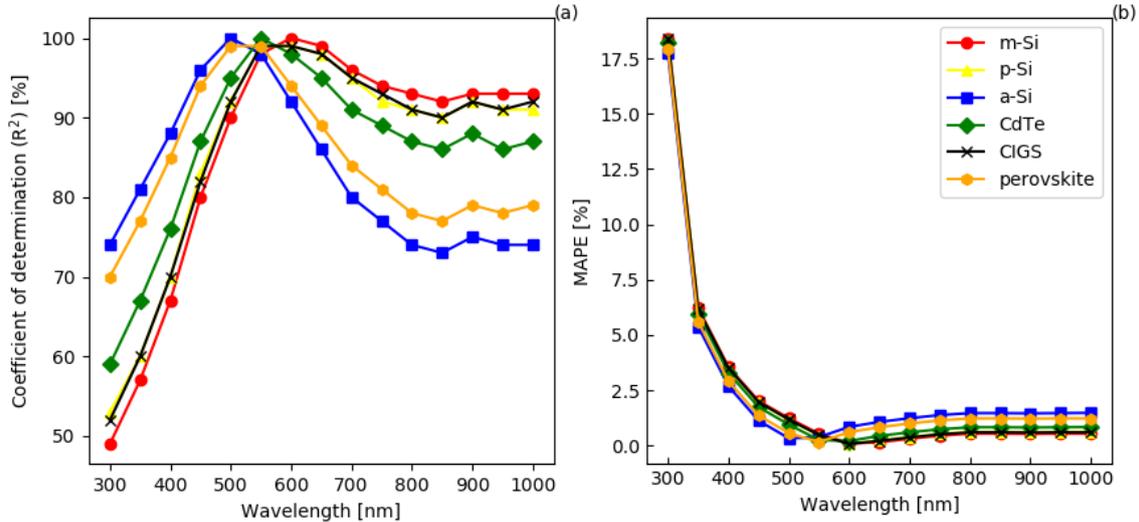

Figure 7. (a) $R^2$ and (b) MAPE obtained when soiling losses for various PV technologies located in Jaén, Spain are estimated using a single hemispherical transmittance wavelength.

The best-performing wavelengths for each material are reported in Table 3. Except for CdTe, the wavelengths that maximize the $R^2$ for a material are those that minimize the MAPE, as well (CdTe's MAPE = 0.21% and $R^2$ = 98.3% at 600 nm). The results show that $R^2$ ≥ 99% and MAPE < 0.35% can be achieved for any PV technology if the transmittance of soiling at a specific wavelength is considered. This suggests that the soiling losses for each material could be ideally predicted by using a single-wavelength measurement with high accuracy. Table 3 suggests that the most appropriate wavelength of each technology can be selected by considering their energy bands: 500 nm for high (a-Si), 550 nm to 600 nm for intermediate (CdTe and perovskite), and 600 nm for low (m-Si, p-Si, and CIGS) energy bandgaps. All materials, except a-Si, show lower MAPE if the transmittance of a single wavelength is used instead of the AST of any of the wavebands investigated in the previous section.

Table 3. Single wavelengths that maximize the coefficient of determination for each PV technology.

|  | Maximum $R^2$ | | |
| --- | --- | --- | --- |
| Material | Wavelength [nm] | $R^2$ (%) | MAPE (%) |
| m-Si | 600 | 99.7 | 0.10 |
| p-Si | 600 | 99.4 | 0.08 |
| a-Si | 500 | 99.7 | 0.33 |
| CdTe | 550 | 99.8 | 0.25 |
| CIGS | 600 | 99.5 | 0.08 |
| perovskite | 550 | 99.0 | 0.16 |

If soiling needs to be determined for more than one PV technology with the same measurement, then it is of interest to find a single wavelength that minimizes the overall error. The coefficients of determination, MAPE, and MPE for each PV technology at the most significant wavelengths found earlier (500, 550, and 600 nm) are reported in Table

4. The transmittance measured at any of the selected wavelengths achieves $R^2 \geq 90\%$ when compared to the soiling ratio of any material. Despite that, the average $R^2$ is lower at the extremes of the selected range. 500 nm favors a-Si and perovskite, but it yields worse predictions for other technology; in contrast, 600 nm maximizes low-energy-band materials, but negatively affects a-Si and perovskite. Moreover, 500 nm shows negative MPE for all the technologies (transmittance systematically lower than soiling ratio) and, in some cases, MAPE is higher than 1%. Therefore, 500 nm seems to be beneficial only if a-Si is investigated. On the other hand, 600 nm should be considered if low-energy bandgap materials are under investigation. Acceptable results are yielded at 600 nm for CdTe, even if 550 nm maximizes its results. Overall, 550 nm can be considered the most convenient if soiling losses need to be determined from one simple wavelength because $R^2$ is equal to or higher than 98% for all the materials.

Table 4. $R^2$, MAPE, and MPE between soiling ratio and transmittance in three spectral regions for the five PV materials considered (for soiling in Jaén, Spain).

| Material | 500 nm | | | 550 nm | | | 600 nm | | |
|---|---|---|---|---|---|---|---|---|---|
| | $R^2$ (%) | MAPE (%) | MPE (%) | $R^2$ (%) | MAPE (%) | MPE (%) | $R^2$ (%) | MAPE (%) | MPE (%) |
| m-Si | 90.3 | 1.24 | -1.24 | 98.3 | 0.54 | -0.54 | 99.7 | 0.10 | -0.08 |
| p-Si | 92.3 | 1.16 | -1.16 | 99.1 | 0.46 | -0.46 | 99.4 | 0.08 | 0.00 |
| a-Si | 99.7 | 0.33 | -0.33 | 97.9 | 0.38 | 0.38 | 91.6 | 0.85 | 0.85 |
| CdTe | 95.4 | 0.95 | -0.95 | 99.8 | 0.25 | -0.25 | 98.3 | 0.21 | 0.21 |
| CIGS | 92.0 | 1.18 | -1.18 | 99.0 | 0.48 | -0.48 | 99.5 | 0.08 | -0.02 |
| perovskite | 98.9 | 0.57 | -0.57 | 99.0 | 0.16 | 0.14 | 94.0 | 0.61 | 0.61 |

The results of this work suggest that soiling detection could be performed by using average waveband or single-wavelength transmittance measurements. The time series of the soiling ratio and of the various indexes here analyzed are shown in Figure 8 for three representative PV materials. Each material has a waveband or wavelength that maximizes the soiling loss prediction, as summarized in Table 5. This can lead to the development of innovative soiling detecting systems, based on transmittance measurements, that might be able to quantify the impact of soiling on different PV technologies.

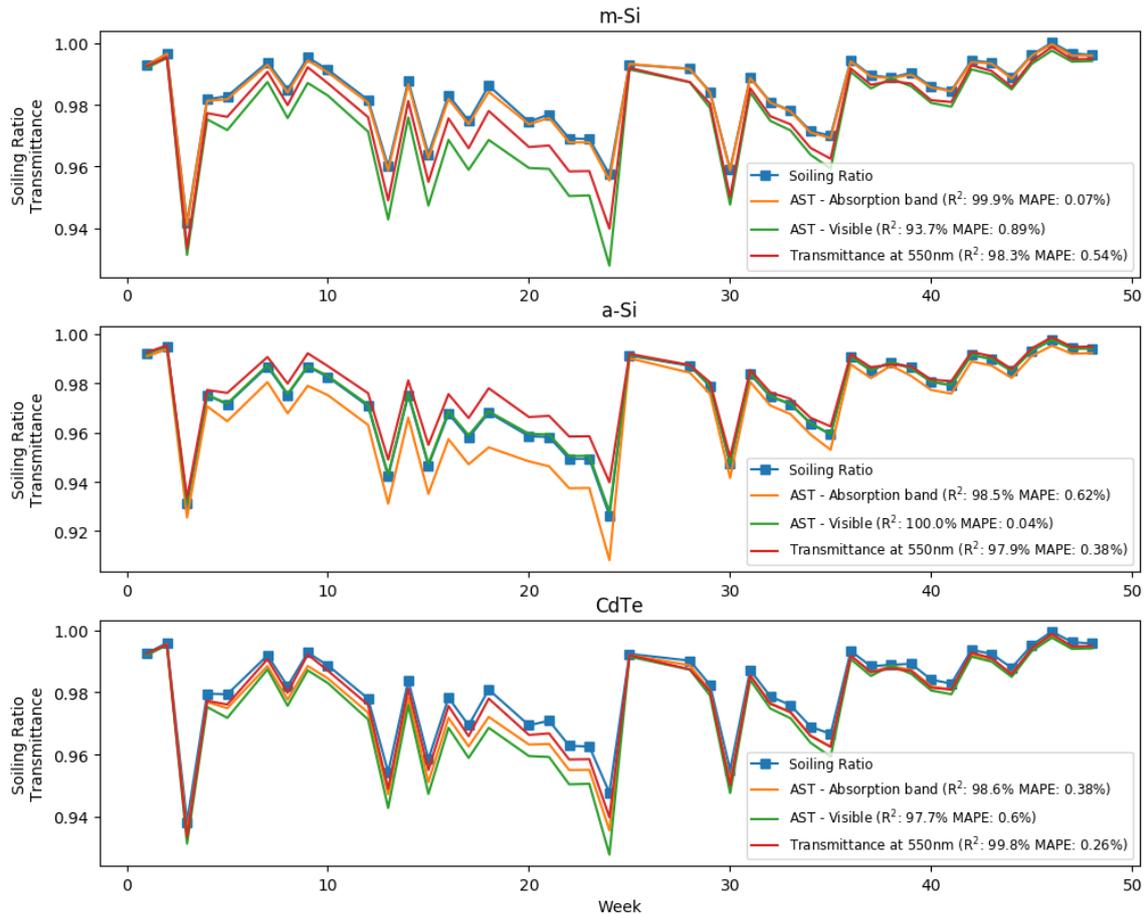

Figure 8. Weekly time series of soiling ratio, AST over the specific material absorption band, and over the visible spectrum and single-wavelength transmittance at 550 nm for m-Si, CdTe, and a-Si.

Table 5. Summary of the best correlations obtained for each material between the soiling ratios and the various parameters investigated.

| Material | Best results |
|---|---|
| m-Si | Transmittance at 600 nm or AST at specific absorption band |
| p-Si | Transmittance at 600 nm |
| a-Si | AST in the visible |
| CdTe | Transmittance at 550 or 600 nm |
| CIGS | AST at specific absorption band, followed by transmittance at 600 nm |
| perovskite | Transmittance at 550 nm, followed by AST in the visible |

*4.3. Discussion*

The experimental investigation here presented was based on a one-year data collection performed in Jaén in Southern Spain. In order to understand the applicability of the presented results to other regions and different soiling conditions, a preliminary analysis of the transmittance at single wavelength, similar to that shown in Section 4.2, has been conducted based on the transmittance losses measured in Golden, Colorado and in San José, California, both in the USA. The coupons are of the same materials and have same size as those used in Jaén, and were horizontally mounted for 6 weeks between January and March in Colorado and for 8 weeks between December and March in California [3]. The weekly hemispherical transmittance between 300 nm and 1100 nm of the coupons in Colorado were measured using a Cary 5000 dual-beam UV-VIS-NIR spectrophotometer equipped with a DRA-2500 integrating sphere. The weekly hemispherical transmittance of the coupons soiled in California was measured with a BWTek iSpec BWS015-Mod fiber optic spectrometer attached to a Spectralon-coated 2 inch diameter Newport integrating sphere. A tungsten lamp allowed for measurements from 410 nm to the NIR, while a phosphor-coated low pressure mercury-vapor lamp (with a peak wavelength near 365 nm) extended the measurements into the UV. The same methodology described in Section 3.3 has been employed, using, in these two cases, the reference ASTM G-173 AM1.5 solar spectral irradiance.

Comparing the results of figures 7 and 9, one sees that similar optimum wavelengths to detect soiling are obtained. Because of the limited number of data points collected at both locations, $R^2$ values (not shown) higher than 97% were found for any wavelength ≥ 400 nm. On the other hand, the analysis of the MAPE returned results similar to those shown earlier. Due to the low transmittance losses recorded in Golden (average transmittance between 0.976 and 1.000), CO, the MAPEs for all the wavelengths are found to be lower than 0.7% (Figure 9a), with values lower than 0.2% for all the materials in the previously mentioned waveband 500 nm to 600 nm. In general, the best wavelengths in Colorado are those between 500 and 700 nm, with a slight red-shift due to the fact that the transmittance losses in the blue spectra are not as high as in conditions of high transmittance loss. Indeed, when a location with higher soiling losses is considered (average transmittance between 0.958 and 0.986), such as San José, CA (Figure 9b), the same correlations as for Jaén are found, with the best wavelengths ranging between 500 nm and 600 nm, where MAPE lower than 0.4% are found.

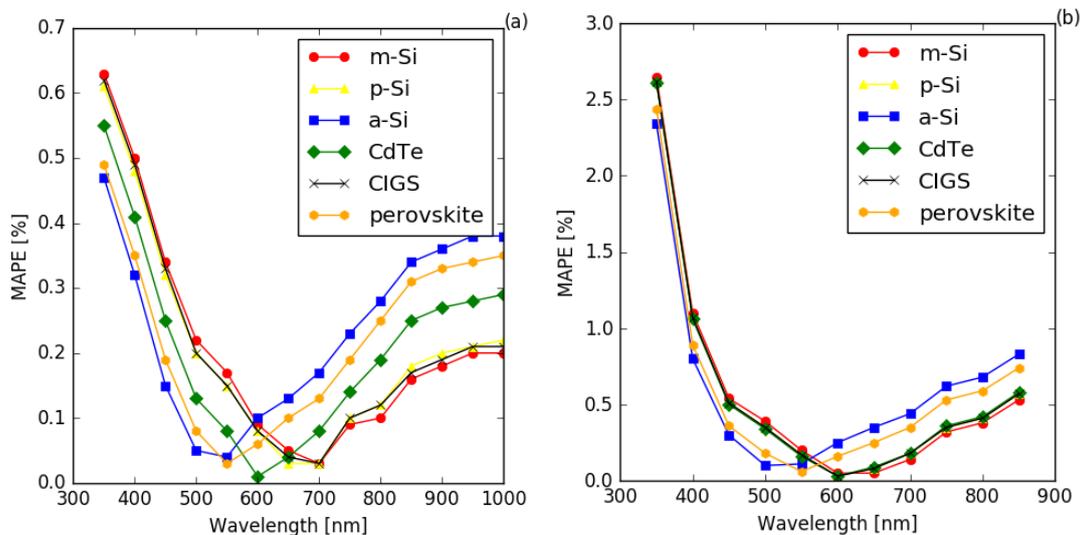

Figure 9. MAPE obtained when soiling losses occurred in Golden, Colorado, (a, left chart) and in San José, California (b, right chart) for various PV technologies are estimated using a single hemispherical transmittance wavelength.

These preliminary results seem to extend and confirm the findings of the investigation conducted in Jaén: it is possible to predict the soiling ratio of different PV modules by measuring the transmittance of soiling at a single wavelength. In light of this, additional studies at different locations are recommended.

## 5. Conclusions

One PV glass coupon was exposed outdoors for 48 weeks in Jaén, a city in southern Spain. The spectral transmittance of the soiling accumulated on the coupon was measured weekly to evaluate the transmittance drop due to soiling and to estimate the losses that this would have caused on PV modules. We investigated the ability to predict soiling losses of different PV materials using only transmittance data. The results show that soiling at a location can, in principle, be estimated by using the transmittance at selected wavelength ranges—or, even at a single wavelength—with high accuracy.

In Jaén, the best estimations are obtained if the transmittance is measured at a wavelength between 500 and 600 nm. Each energy-band category shows a range in which the results are optimized: 500 nm for high (a-Si), 550 nm to 600 nm for intermediate (CdTe and perovskite), and 600 nm for low (m-Si, p-Si and CIGS) energy bandgaps. Alternatively, the average spectral transmittance over the specific material absorption band returns the best soiling estimates compared to the average transmittance of the spectral regions for all materials, except for amorphous silicon and perovskite cells. Among the three regions of the solar irradiance, the best results are obtained for the visible band ($R^2 \geq 94\%$ for all the materials), even if this introduces a systematic offset in calculating the soiling ratio for low-energy-bandgap materials (as shown for the m-Si time series in Figure 8).

The findings of this work results can lead to the development of innovative spectral soiling-detector devices. The data collection used for this analysis took place over 48 weeks at only one location that, over a one-year period, is exposed to various types of soiling (Saharan dust, olive tree pollen and smoke, and urban particulate matter). Even so, similar investigations should be replicated at different locations. Preliminary analogous investigations, conducted for 6 to 8 weeks at sites exhibiting different soiling conditions, confirmed that the best wavelengths for the estimation of soiling are found to range between 500 nm and 700 nm, with a slight red-shift in the results for conditions of low-soiling. Since the transmittance in this study was measured at zero angle of incidence, whereas for PV modules deployed in the field, the transmittance of soiling varies daily and hourly (depending on the angle of incidence), further studies should be conducted to consider these effects and to confirm the findings of the present study.

**Acknowledgments**

This work was conceived and partially funded as part of the "Global investigation on the spectral effects of soiling losses" project, financed under the EPSRC SUPERGEN SuperSolar Hub's "International and industrial engagement fund."

This research was partially funded by the Universidad de Jaén (UJA) and Caja Rural de Jaén, grant number UJA2015/07/01.

This work was authored in part by Alliance for Sustainable Energy, LLC, the manager and operator of the National Renewable Energy Laboratory for the U.S. Department of Energy (DOE) under Contract No. DE-AC36-08GO28308. Funding provided by the U.S. Department of Energy's Office of Energy Efficiency and Renewable Energy (EERE) under Solar Energy Technologies Office (SETO) Agreement Number 30311. The views expressed in the article do not necessarily represent the views of the DOE or the U.S. Government. The U.S. Government retains and the publisher, by accepting the article for publication, acknowledges that the U.S. Government retains a nonexclusive, paid-up, irrevocable, worldwide license to publish or reproduce the published form of this work, or allow others to do so, for U.S. Government purposes.

This study is partially based upon work from COST Action PEARL PV (CA16235), supported by COST (European Cooperation in Science and Technology). COST (European Cooperation in Science and Technology) is a funding agency for research and innovation networks. Our Actions help connect research initiatives across Europe and enable scientists to grow their ideas by sharing them with their peers. This boosts their research, career and innovation, see www.cost.eu.

Part of this work was funded through the European Union's Horizon 2020 research and innovation programme under the NoSoilPV project (Marie Skłodowska-Curie grant agreement No. 793120).